\newcommand\araa{{ARA\&A\,}}%
\newcommand\aap{{A\&A\,}}%
\newcommand\mnras{{MNRAS\,}}%
\newcommand\assl{{Astrophysics and Space Science Library\,}}

\documentclass[graybox, natbib]{svmult}

\usepackage{natbib}       

\usepackage{mathptmx}       
\usepackage{helvet}         
\usepackage{courier}        
\usepackage{makeidx}         
\usepackage{graphicx}        
\usepackage{multicol}        
\usepackage[bottom]{footmisc}

\makeindex             

\begin{document}

\title*{Mapping the stellar populations of the Milky Way with Gaia}
\titlerunning{The MW stellar populations with Gaia}
\author{Carla Cacciari}
\institute{Carla Cacciari \at INAF - Osservatorio Astronomico, Via Ranzani 1, Bologna 
\email{carla.cacciari@oabo.inaf.it}
}
%
%
\maketitle

\abstract{Gaia will be ESA's milestone astrometric mission, and is due for launch 
at the end of 2013. Gaia will repeatedly map the whole sky measuring about one billion 
sources to V=20-22 mag. Its data products will be $\mu$as accuracy astrometry, 
optical spectrophotometry and medium resolution spectroscopy. 
A description of the Gaia space mission and its characteristics and 
performance is given. The expected impact on Galactic stellar population studies 
is discussed, with particular attention to the sources of interest for CoRoT and 
{\it Kepler}.  
}

\section{Gaia}\label{s:gaia}           

Gaia is a major ESA mission with astrometric, photometric and spectroscopic 
capabilities. It is currently scheduled for launch in December 2013 from Kourou, to 
be placed at the Lagrangian point L2, 1.5 million km from the Earth in the 
direction opposite the Sun, for a planned lifetime of 5 years.\footnote{For a 
detailed description of the satellite and system, payload and telescope, 
functioning and operations see \url{http://www.rssd.esa.int/index.php?project=gaia  
Information sheets}.}

Gaia represents the natural continuation and a huge improvement with respect  
to the Hipparcos mission: it will extend the V magnitude limit from 12 
to about 20-22 (for blue and red objects respectively), observe a factor 10$^4$ 
more sources (including objects such as galaxies and quasars unobservable by 
Hipparcos), reach a factor $\sim$100 better astrometric accuracy, and provide 
spectrophotometric information for all of the observed objects, as well as 
spectroscopy for a large fraction of them.  

These characteristics are summarized in Table \ref{t:1}, and are described in 
some more detail in the following.    

\begin{table}
\caption{From Hipparcos to Gaia.}
\label{t:1}       
%
%
\begin{tabular}{p{2.5cm}p{2.5cm}p{6.3cm}}
\hline\noalign{\smallskip}
 & Hipparcos$^a$ &  Gaia$^b$  \\
\noalign{\smallskip}\svhline\noalign{\smallskip}
 Magnitude limit   & V$_{lim}$ = 12 & V$_{lim}$ = 20-22 (blue-red sources, respectively) \\
 N. of objects     & 1.2$\times$10$^5$  & $\ge$10$^9$ (2.5$\times$10$^7$ to V=15, 2.5$\times$10$^8$ to V=18) \\
 Quasars           & none & $\sim$5$\times$10$^5$ \\
 Galaxies          & none & $\sim$ 10$^6$-10$^7$ \\
 Astrom. accuracy  & $\sim$1 mas & $\sim$7-10$\mu$as at V$\le$12 \\
                   &             & 10-25 $\mu$as at V=15, 100-300 $\mu$as at V=20 \\
 Broad-band phot.  & 2 (B,V) & 3 (to V$_{lim}$) + 1 (to V=17) \\
 Spectrophtometry  & none & 2 bands (B/R) to V$_{lim}$ \\
 Spectroscopy (CaT)& none & 1-15 km/s to V=16-17 \\
 Obs. programme    & pre-selected targets & all-sky complete and unbiased \\
\noalign{\smallskip}\hline\noalign{\smallskip}
\end{tabular}
$^a$  Final Catalogue: \cite{Perryman97}; New Reduction of the Raw Data: \cite{FvL07}\\
$^b$ Expected Final Catalogue: 2020-22
\end{table}

\subsection{Satellite, payload, instruments}      

\begin{figure}[h]
\sidecaption
\includegraphics[scale=0.45]{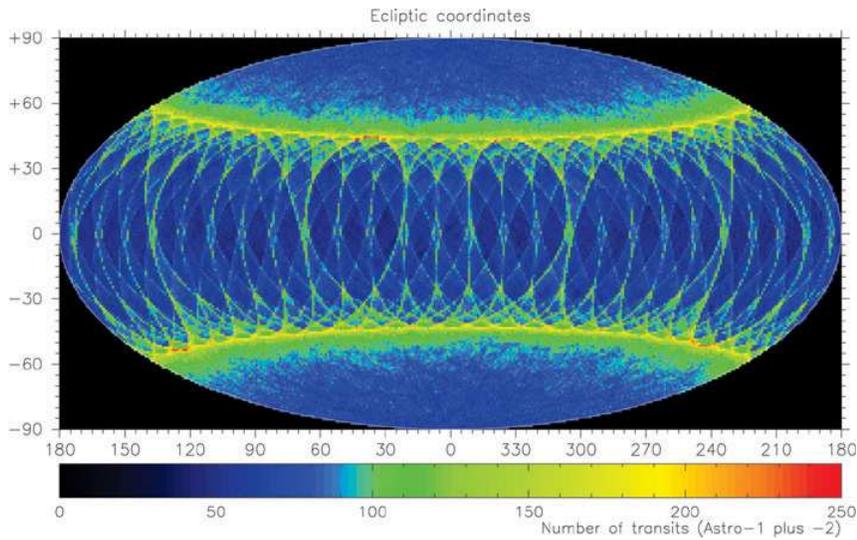}
%
%
\caption{Dependence of the end-of-mission number of focal plane transits
on position on the sky. Shown is an all-sky equal-area Hammer
projection in ecliptic coordinates.  The maximum number of transits
will occur in a $\sim$ 10-deg wide strip around ecliptic latitudes
+/-- 45 deg. }
\label{f:scanlaw}       
\end{figure}

The satellite spins around its axis, which is oriented 45-deg away from  
the Sun, with a period of 6 hr, and   
the spin axis has a precession motion around the solar direction with 
a period of 63 days.  
The combination of these two motions results in the scanning law that 
allows the entire sky to be observed on average 70 times 
over the 5 yr mission lifetime (see the transit map in Fig. \ref{f:scanlaw}). 

The payload  is a toroidal structure holding two primary 1.45m$\times$0.50m 
rectangular mirrors (field of view FoV = 1.7-deg$\times$0.6-deg)  whose lines 
of sight are separated by an angle of 106.5-deg (Basic Angle, BA). 
The BA needs to be known with extremely high precision to ensure 
Gaia's expected astrometric accuracy, and therefore a BA 
monitoring system is hosted on the payload, as well as all the optical 
components which allow to superpose the FoVs  of the two  mirrors and 
combine them on the focal plane. 

The focal plane contains several arrays of 4.5K$\times$2.0K CCDs: \\
{\bf i)} the sky mapper (SM), 2$\times$7 CCDs for detection and confirmation of 
source transit; \\
{\bf ii)} the astrometric field (AF), 9$\times$7 CCDs corresponding to 
40$\times$40 arcmin, for astrometric measurements and white light 
(G-band, 330-1050 nm) photometry; \\
{\bf iii)} the blue (BP) and red (RP) photometers, 2$\times$7 CCDs for low 
resolution (R$<$100) slitless prism spectro-photometry in the ranges 330-680 nm 
and 640-1050 nm, respectively. From the spectra the G$_{BP}$ and G$_{RP}$ 
integrated  magnitudes are derived.\\
{\bf iv)} the radial velocity spectrometer (RVS), 3$\times$4 CCDs for slitless 
spectroscopy 
at the Ca II triplet (847-870 nm) with R$\sim$11,000.\\
Measurements are made in time-delayed-integration mode, reading the CCD
at the same speed as the source trails across the focal plane, i.e. 
60 arcsec/sec, corresponding to a crossing/reading time of 4.4 sec per CCD.

%

\subsection{Astrometry: measuring principles}\label{ss:prin}

The mission is designed to perform global (wide field) astrometry as
opposed to local (narrow field) astrometry. In local astrometry, the
star positions can only be measured with respect to neighbouring stars
in the same field. Even with a very accurate instrument the propagation
of errors is prohibitive when making a sky survey. The principle of
global astrometry is to link stars with large angular distances in a
network where each star is connected to a large number of other stars
in every direction.

Global astrometry requires the simultaneous observation of two fields
of view in which the star positions are measured and compared. 
This is provided by the two lines of sight of the primary mirrors.
Then, like with Hipparcos, the two images are combined, slightly spaced, 
on a unique focal plane assembly. 
Objects are matched in successive scans, attitude and
calibration parameters are updated, and object positions are solved and 
fed back into the system. This procedure is iterated as more scans are added. 
In this way the system is self-calibrating by the use of isolated non variable 
point sources, which will form a sufficiently large body of reference objects for 
most calibration purposes, including the optical definition of the International 
Celestial Reference System (ICRS) by observing about half a million QSOs. 

Therefore Gaia's astrometry will be not only unprecedently accurate as far 
as internal rms errors are concerned, because derived from an all-sky 
solution, but also unprecedently precise in {\em absolute} values, because 
obtained with {\em direct reference to the ICRS}.

\subsection{Expected performance}\label{ss:perf}

$Astrometry$  ~~Astrometric errors are dominated by photon statistics. Sources
at V$\sim$6 mag represent the bright magnitude limit for Gaia  observations,
as saturation sets in at that level. The predicted sky-averaged end-of-mission 
standard errors on the parallax are summarized in Table 
\ref{t:perf}\footnote{See \url{http://www.rssd.esa.int/index.php?project=gaia} 
Science Performance,  April 2011 update.}. 
We note that the standard errors on position and proper motion are 
about 0.74 and 0.53 of those on parallax, respectively.

$Photometry$ ~~  Gaia's photometric data include the integrated white light 
(G-band) from the AF, and the BP/RP prism spectra from which the G$_{BP}$ and 
G$_{RP}$ integrated magnitudes are derived. The expected end-of-mission errors 
are shown in Table \ref{t:perf}. 

From the BP/RP spectral energy distributions it will be possible to
estimate astrophysical parameters using pattern recognition techniques
(\cite{Bailer-Jones10}).  For example, one may expect to obtain: 
i) T$_{eff}$ to $\le$5\% (15\%) for a wide range of spectral types brighter (fainter) 
than V=16; 
ii) log$g$~ to 0.2-0.3 dex  (0.2-0.5 dex) for hot (SpT $\le$ A) stars brighter (fainter) 
than V=16; 
iii)  $[Fe/H]$~ to $\sim$ 0.2-0.4 dex (0.5-0.7 dex) down to [Fe/H]=--2.0 for cool stars 
(SpT $>$ F) brighter (fainter) than V=16; 
iv)  A$_V$ to $\sim$ 0.05-0.2 mag (0.05-0.3 mag)  for hot stars brighter (fainter) than V=16. 

Ranges in errors reflect the influence of spectral type and metallicity. It is also 
to be noted that at V=15 the degeneracy between  T$_{eff}$ and A$_V$ amounts 
to about 3-4\% and 0.1-0.2 mag respectively.

$Spectroscopy$~~ The RVS provides the third component of the space velocity for 
red (blue) sources down to about magnitude 17 (16).
Radial velocities are the main product of the RVS, with typical end-of-mission 
errors as shown in Table \ref{t:perf}. 
For sources brighter than $\sim$ 14 mag the RVS spectra will provide information 
also on rotation and chemistry, and combined with the prism BP/RP spectra 
will allow us to obtain more detailed and accurate astrophysical parameters.

\begin{table}
\caption{End-of-mission expected standard errors of astrometric, photometric and spectroscopic data 
as a function of Johnson V magnitude for three unreddened reference spectral types. 
  Left: sky-averaged parallax errors (in units of $\mu$as) for B1V, G2V, and M6V. 
  Middle: photometric errors in the G-G$_{BP}$-G$_{RP}$ bands, in units of milli-magnitude,  
  for  B1V, G2V, and M6V. 
  Right: radial velocity errors (in units of km/s) for  A0V, G5V and K4V. }
\label{t:perf}
\centering
\begin{tabular}{c | ccc | ccc | ccc}
\hline
& & & & &  & & & &  \\
V &     & $\sigma_{\pi}$ ($\mu$as) &  &     & $\sigma_{phot}$ (mmag) &  & & $\sigma_{RV}$ (km/s)  \\
~~~(mag)~~~  & ~~B1V & G2V  & M6V~~ & B1V & G2V  & M6V  & ~~A0V & G5V  & K4V   \\ 
\hline
& & & & &  & & & &  \\
6-12   & 7   & 7  & 7   & 1-4-4  & 1-4-4  & 1-4-4     & 1-2   & 1   & 1   \\
13     &  11   &  10.5    &  7.5     & 1-4-4  & 1-4-4  & 1-4-4     & 3     & 1   & 1   \\
15     & 27     & 26    & 10      & 1-4-5  & 1-4-4  & 1-6-4     & 16    & 3   & 2 \\ 
16     & 41     &  41   &  15    & 1-4-5  & 1-5-5  & 1-9-4     &       & 7-8 & 4   \\
17     &  70    &   66  &   23   & 2-5-7  & 2-5-5  & 2-20-5    &       & 20  & 10 \\
18     &  110    &   107  &  40    & 2-7-14 & 2-9-8  & 2-49-5    &       &     &     \\ 
20     & 340    & 333   & 100    & ~~3-29-83~~ & 3-43-43 & ~~3-301-17~~ &       &     &  \\
\hline
\end{tabular}
\end{table}

\section{Science with Gaia}\label{s:science}  

``The primary objective of Gaia is the Galaxy: to observe the physical
characteristics, kinematics and distribution of stars over a large fraction
of its volume, with the goal of achieving a full understanding of the MW
dynamics and structure, and consequently its formation and history.'' 
(Concept and Technology Study Report, ESA-SCI-2000-4). \\
Gaia will provide a complete census of all Galactic stellar populations 
down to 20th magnitude.  Based on the Besan{\c c}on Galaxy model 
(\citealt{Robin03}, \citealt{Robin04} ) Gaia is expected to measure more than 10$^9$  stars 
belonging to the thin and thick disk, the bulge and the spheroid. 
Binaries, variable stars and rare (i.e. fast-evolving) stellar types will 
be well sampled, as well
as special objects such as Solar System bodies ($\sim 10^5$),
extra-solar planets ($\sim 2\times10^4$), WDs ($\sim 2\times10^5$),
BDs ($\sim 5\times10^4$).

One billion stars in 5-D (6-D if the radial velocity is available, and
up to 9-D if the astrophysical parameters are know as well) will allow
us to derive the spatial and dynamical structure of the Milky Way, its
formation and chemical history (e.g. by detecting evidence of
accretion/merging events), and the star formation history throughout
the Galaxy.  The huge and accurate database will provide a powerful
testbench for stellar structure and evolution models.  
The superb astrometric and photometric accuracy will allow us 
to obtain proper-motion-cleaned  Hertzprung-Russell diagrams 
throughout the Galaxy, and hence complete characterization and dating 
of all spectral types and Galactic stellar populations.  
The dark matter distribution will be mapped by the distribution and rate 
of microlensing events. 
The cosmic distance scale will be accurately defined on a reliable ground  
thanks to the distance (i.e. luminosity) calibration of the 
primary standard candles, RR Lyraes and Cepheids.

\subsection{MW stellar population studies with Gaia: a few examples}\label{ss:pops}

\noindent $The~ Bulge: ~~\sim 1.7\times10^8 stars$~~~ 
Our knowledge of the Galactic bulge has greatly improved in the last decade(s), and 
presently the structure of the bar is constrained from modelling of gas dynamics, 
stellar surface brightness and stellar dynamics. 
However many questions remain open, for example (just to quote a few) on the formation 
mechanism (single enrichment event or merging from chemically distinct subcomponents?), 
on the chemical evolution timescale (less than 1 Gyr or more extended?), on the 
presence of chemodynamical subpopulations, on the relation with other Galactic 
populations, especially the inner disk (see \citealt{Rich13} for a review). 

Gaia will make a major contribution to the solution of these problems by 
measuring accurate distances and proper motions of several millions of stars, as well as 
a huge number of radial velocities (to V=17 mag), especially in bulge fields at 
$b < -6^{\circ}$ where the X-shaped structure is important. The northern bulge will 
also be observable, because red clump stars can be reached by Gaia even with 
$\sim$3-4 mag extinction. Simulations show that, at the reference distance of 
8 kpc,  a typical tracer such as a red clump star 
(M0III, M$_V$=--1 mag) dimmed by 4 mag extinction would have V=17.5. 
Gaia will measure its parallax with an rms error of $\sim$50 $\mu as$ and  
proper motion to $\sim$1 km/s, as well as the radial velocity to $\le$15 km/s,  
and obtain useful information on its astrophysical parameters (and hence age 
and chemical properties).     
Complementary high-dispersion spectroscopy, e.g. by the Gaia-ESO Survey (GES), 
HERMES and 4MOST, will provide more detailed information on chemistry and 
kinematics. \\

\noindent $The~ Disk(s): ~~\ge 10^9 stars$~~~
In recent years several photometry and spectroscopy surveys have greatly increased 
the number of stars with good distances, radial and transverse velocities, and
abundance estimates. However, an enormous amount of practical and conceptual work 
needs to be done to answer the many questions still open in this field (see 
\cite{Rix-Bovy13} for a review).

Based on current simulations, the effective volume that Gaia will explore will be 
limited to only a quadrant of the Galactic disk, because of dust extinction and image 
crowding. 
Assuming as a typical tracer a K3III star (M$_V$=0 mag) dimmed by 2 mag extinction, 
the disk can be mapped as far as 10 kpc with individual distance errors $\le$50 $\mu as$, 
proper motion errors $\le$1.5 km/s, radial velocity errors $\le$10 km/s, and with additional  
information on astrophysical parameters and ages. 
The huge number and high accuracy of these data will provide a major breakthrough 
in the understanding of the many aspects related to the disk formation, structure 
and evolution (see K. Freeman's contribution, this meeting).   \\

\noindent $The~ Halo: ~~\sim 2\times10^7 stars$~~~
Typical tracers of the field halo population, which have been used in several studies,  
are red giants (K3III, M$_V$=-1), HB stars (A5III, M$_V$=+0.5), and MS-TO stars 
(G2V, M$_V$=+4.5). 
With the former two stellar types Gaia will map the inner halo as 
far as 10 kpc with proper motion errors $\le$1 km s$^{-1}$, and the 
outer halo as far as 30 kpc with proper motion errors of $\sim$ 3-7 km s$^{-1}$, 
respectively. The much fainter (but more numerous) MS-TO stars can be used to map 
the inner halo as far as 4 kpc with proper motion errors $\le$1 km s$^{-1}$ (as far 
as 10 kpc to 6-7 km s$^{-1}$). 
By these in-situ measurements it will be possible to settle questions such as 
the inner/outer halo dichotomy, their origins and mechanisms of formation (if 
different) and hence the merger history of the Galaxy, and derive the 
gravitational potential of the Milky Way's dark matter halo. The synergy with 
LSST will be especially fruitful to extend these results at fainter magnitudes 
(see \citealt{Ivezic12} for a review). \\

\section{CoRoT and {\it Kepler} targets}\label{s:corkep} 

The study of the dynamical and chemical evolution of stellar populations
in the Galaxy requires accurate data on kinematics (velocities), chemical 
properties (abundances) and ages (distances, astrophysical parameters) for 
a significant fraction of MW stars. 
Gaia, in synergy with the large ongoing or forthcoming photometric and spectroscopic 
surveys, will provide enormous amounts of such data in the next decade. 
Accurate ages for individual field stars, however, are quite difficult to acquire, 
and asteroseismology can make a fundamental contribution by estimating ages,  
e.g. for individual red giants.  These stars are bright and allow us to probe the 
evolution of populations across the whole Galaxy as far as its more distant 
parts. 

During their searching  campaigns for exoplanets around late type (mostly F-M) dwarf 
stars, CoRoT and {\it Kepler} found thousands of red giants with solar-like 
oscillations, which could be analyzed with asteroseismology thechniques to derive 
their physical parameters, distances and ages.  
As an example, we consider the work by \cite{Miglio13} 
who analyzed 
about 2000 red giants with solar-like oscillations in two exofields of the CoRoT 
survey extending about 8-10 kpc each, on opposite directions with respect to the Sun.  
The stars are selected to be brighter than R=16. For our simulations of Gaia 
observations we have assumed as templates the spectral types G8III, K3III and M0III. 
For the sake of completeness, we have performed simulations also for two 
template red dwarfs, i.e. F6V and G2V, which are targets of the CoRoT and {\it Kepler} 
surveys as well.  

The characteristics of these stars, and the expected astrometric and kinematic performance 
of Gaia at the adopted R$_{lim}$=16 are summarized in 
Table \ref{t:corot}\footnote{Astrometric simulations are based on de Bruijne (2009), 
GAIA-CA-TN-ESA-JDB-055.}. 
The Gaia photometric errors at these levels of magnitude are a few mmag   
(see Table \ref{t:perf}). 
As mentioned in Sect. \ref{ss:perf}, for stars brighter than $\sim$ 16 mag the 
spectrophotometric data are expected to give good information on stellar 
astrophysical parameters such as temperature, gravity, metallicity and  
reddening; somewhat less accurate, but still useful estimates of these 
parameters can be obtained down to $\sim$ 18 mag.

\begin{table}
\caption{CoRoT red stars with solar-like oscillations: expected end-of-mission errors on parallax, 
transverse velocity (from proper motions) and radial velocity (from the RVS) from Gaia measures, 
at the magnitude limit R$_{lim}$=16 assuming zero reddening. 
 }
\label{t:corot}
\centering
\begin{tabular}{cccccccc}
\hline
& & & & &  & &  \\
Sp.~Type~~~ & ~~~~V-R~~~~&~~~~V~~~~&~~~~M$_V$~~~~&~~~~Dist - Par~~~~& ~~~~$\sigma_{\pi}$~~~~ & ~~~p.m. $\sigma_{TV}$~~~ & ~~RVS $\sigma_{RV}$~~   \\
            &   (mag)    & (mag)   & (mag)       &(kpc) - ($\mu$as) &  ($\mu$as)             &   (km/s)                 &   (km/s)   \\ 
\hline
& & & & &  & &  \\
G8III     & 0.50   & 16.50  & +0.6   &   15 - 66  &  40     & 1.5     &	 11  \\ 
K3III     & 0.64   & 16.60  & +0.1   &   20 - 50  &  40	    & 2.0     &   5  \\
M0III     & 0.88   & 16.90  & -0.4   &   29 - 35  &  39     & 2.9     &   7  \\
& & & & &  & &  \\
F6V       & 0.29   & 16.30  & +3.5   &  3.6 - 275 &  39     & 0.3-0.4 &  15  \\
G2V       & 0.37   & 16.40  & +4.8   &  2.1 - 480 &  39     & 0.2     &  12  \\
\hline
\end{tabular}
\end{table}

\begin{figure}[h]
\sidecaption
\includegraphics[scale=0.65]{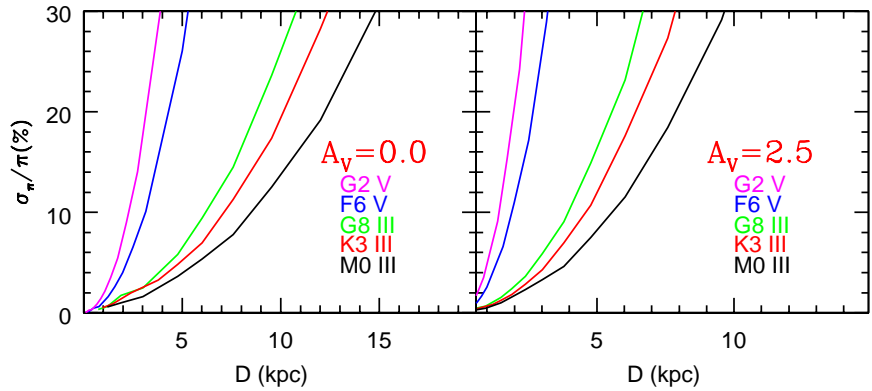}
\caption{The lines show Gaia's expected accuracy on parallax measures as a function of distance, 
for the same spectral types listed in Table \ref{t:corot}.   }
\label{f:corkep}       
\end{figure}

To complement the information given in Table \ref{t:corot}, we plot in 
Fig. \ref{f:corkep} the detailed behaviour of the expected astrometric 
accuracy (i.e. percent error on parallax) as a function of distance, 
which shows the range of distances that can be reached by the various 
spectral types to $\le$30\%, for two values of reddening. \\

The results of these simulations indicate that there is an area of overlap 
where the combined use of Gaia's data and asteroseismology thechniques can be 
very fruitful. 
On the one hand, Gaia will provide a complete census and very accurate multi-fold 
information 
for the nearest and brightest stars accessible to asteroseismology, 
and hence facilitate the selection of targets and initial input parameters.  
On the other hand, the cross-verification of Gaia's results by the very detailed 
and independent asteroseismology thechniques, even if applicable only to a small 
subsample of stars, will help understand in finer detail Gaia's results and 
calibrate them on physical grounds. Possible discrepancies, if any, only promise 
to open the path to deeper understanding.

%

\begin{acknowledgement}
The support by the INAF (Istituto Nazionale di Astrofisica) and the ASI 
(Agenzia Spaziale Italiana) under contracts I/037/08/0 and
I/058/10/0 dedicated to the Gaia mission is gratefully acknowledged.
\end{acknowledgement}
%

%



\begin{thebibliography}{9}
\expandafter\ifx\csname natexlab\endcsname\relax\def\natexlab#1{#1}\fi

\bibitem[{{Bailer-Jones}(2010)}]{Bailer-Jones10}
{Bailer-Jones}, C. 2010, \mnras, 403, 96

\bibitem[{{Ivezi\'c} {et~al.}(2012){Ivezi\'c}, {Beers}, \&
  {Juri\'c}}]{Ivezic12}
{Ivezi\'c}, Z., {Beers}, T., \& {Juri\'c}, M. 2012, \araa, 50, 251

\bibitem[{{Miglio} {et~al.}(2013){Miglio}, {Chiappini}, {Morel}, {Barbieri},
  {Chaplin}, {Girardi}, {Montalb{\'a}n}, {Valentini}, {Mosser}, {Baudin},
  {Casagrande}, {Fossati}, {Aguirre}, \& {Baglin}}]{Miglio13}
{Miglio}, A., {Chiappini}, C., {Morel}, T., {et~al.} 2013, \mnras, 429, 423

\bibitem[{{Perryman}(1997)}]{Perryman97}
{Perryman}, M. 1997, ESA-SP, 402

\bibitem[{{Rich}(2013)}]{Rich13}
{Rich}, R. 2013, in \textit{Planets, Stars and Stellar Systems: Galactic
  Structure and Stellar Populations} eds. T.D. Oswalt, G. Gilmore, Springer, 5

\bibitem[{{Rix} \& {Bovy}(2013)}]{Rix-Bovy13}
{Rix}, H. \& {Bovy}, J. 2013, arXiv:1301.3168

\bibitem[{{Robin} {et~al.}(2003){Robin}, {Reyl{\'e}}, {Derri{\`e}re}, \&
  {Picaud}}]{Robin03}
{Robin}, A.~C., {Reyl{\'e}}, C., {Derri{\`e}re}, S., \& {Picaud}, S. 2003,
  \aap, 409, 523

\bibitem[{{Robin} {et~al.}(2004){Robin}, {Reyl{\'e}}, {Derri{\`e}re}, \&
  {Picaud}}]{Robin04}
{Robin}, A.~C., {Reyl{\'e}}, C., {Derri{\`e}re}, S., \& {Picaud}, S. 2004,
  \aap, 416, 157

\bibitem[{{van Leeuwen}(2007)}]{FvL07}
{van Leeuwen}, F. 2007, \assl, Springer, 350

\end{thebibliography}

%
%
\end{document}